\documentclass[a4paper,10pt]{scrartcl}

\usepackage{amsmath}
\usepackage{amsfonts}
\usepackage{amssymb}
\usepackage{graphicx}
\usepackage[english]{babel}
\usepackage{wasysym}
\usepackage{units}

\newcommand{\p}{\partial}
\newcommand{\reff}[1]{(\ref{#1})}
\newcommand{\vs}[1]{\vspace{#1mm}}
\newcommand{\vsO}{\vspace{.1cm}\hfill\\}
\newcommand{\vsT}{\vspace{.2cm}\hfill\\}

\title{Study of the Quasi-isotropic Solution near the Cosmological Singularity in Presence of Bulk-Viscosity}

\author{Nakia Carlevaro$^{\;a,\;b}$ and Giovanni Montani$^{\;b,\;c,\;d,\;e}$\vsT
\emph{\footnotesize $^a$Department of Physics, Polo Scientifico -- Universit\`a degli Studi di Firenze,}\vs{-2.5}\\
\emph{\footnotesize INFN -- Section of Florence, Via G. Sansone, 1 (50019), Sesto Fiorentino (FI), Italy}\\
\emph{\footnotesize $^b$ICRA -- International Center for Relativistic Astrophysics,}\vs{-2.5}\\
\emph{\footnotesize c/o Dep. of Physics - ``Sapienza'' Universit\`a di Roma}\\
\emph{\footnotesize $^c$ Department of Physics - ``Sapienza'' Universit\`a di Roma, Piazza A. Moro, 5 (00185), Rome, Italy}\\
\emph{\footnotesize $^d$ENEA -- C.R. Frascati (Department F.P.N.), Via Enrico Fermi, 45 (00044), Frascati (Rome), Italy}\\
\emph{\footnotesize $^{e}$ ICRANet -- C. C. Pescara, Piazzale della Repubblica, 10 (65100), Pescara, Italy}\vsO
{\footnotesize\ttfamily nakia.carlevaro@icra.it\quad montani@icra.it}
}
\date{}
\begin{document}
\maketitle

%
\begin{abstract} \textbf{Abstract:}
We analyze the dynamical behavior of a quasi-isotropic Universe in the presence of a cosmological fluid endowed with \emph{bulk viscosity}. We express the viscosity coefficient as a power-law of the fluid energy density: $\zeta=\zeta_0\,\epsilon^{s}$. Then we fix $s=\nicefrac{1}{2}$ as the only case in which viscosity plays a significant role in the singularity physics but does not dominate the Universe dynamics (as requested by its microscopic perturbative origin). The parameter $\zeta_0$ is left free to define the intensity of the viscous effects.

Following the spirit of the work by E.M. Lifshitz and I.M. Khalatnikov on the quasi-isotropic solution, we analyze both Einstein and hydrodynamic equations up to first and second order in time. As a result, we get a power-law solution existing only in correspondence to a restricted domain of $\zeta_0$.
\end{abstract}

\section{Introduction}
Near the cosmological singularity, the isotropic nature of the Universe corresponds to a class of solutions of the Einstein equations containing three physically arbitrary functions of the space coordinates. In the case of a radiation-dominated Universe, such a class was found by E.M. Lifshitz and I.M. Khalatnikov in 1963 \cite{lk63} and then generalized to an arbitrary fluid state equation in \cite{kks02}. Earlier extensions of this quasi-isotropic scheme were provided in \cite{mont99,mont00,im03,td94}, where different evolutionary stages of the Universe are characterized.

The original work by Lifshitz and Khalatnikov treated the quasi-isotropic model as a Taylor expansion of the 3-metric tensor in powers of the synchronous time. However, further investigation outlined the necessity of treating generic power-law components of the 3-metric. We fix our attention on the relevance of dealing with bulk viscous properties of the cosmological fluid approaching the Big-Bang singularity. In fact, \emph{bulk viscosity} is a phenomenological issue inherent to the difficulty for a thermodynamical system to follow the equilibrium configuration. Since asymptotically near the Big-Bang the volume of the Universe has a very fast time variation, we naturally expect the bulk viscous effects arise in the dynamics. It is worth noting that such viscous contributions can not dominate the Universe evolution because of their thermodynamical perturbative origin; nevertheless, we are interested in those regimes where such effects are not at all negligible.

General analyses of the Universe behavior in presence of bulk viscosity were given in \cite{bk76,bk77,bnk79} in which the bulk viscosity coefficient $\zeta$ is described by a power-law of the energy density of the fluid, \emph{i.e.}, $\zeta=\zeta_0\,\epsilon^{s}$ ($\zeta_0,\,s=const$). As far as this phenomenological \emph{ansatz} is referred to the early isotropic Universe, it is easy to realize that the choice $s=\nicefrac{1}{2}$ prevents dominating viscous effects. On the other hand, simple considerations, as well as the analysis presented in the works by J.D. Barrow \cite{b87,b88,b90}, indicate that the case $s<\nicefrac{1}{2}$ leads to negligible contributions of the viscosity to the asymptotic regime towards the Big-Bang. As a consequence, in studying the singularity physics, the most appropriate form of the power-law is $\zeta=\zeta_0\,\sqrt{\epsilon}$.

In \cite{nkm05}, the stability of a flat Friedmann - Robertson - Walker (FRW) Universe was investigated including bulk viscosity in the cosmological fluid dynamics (see also \cite{paddy}). As a result, it was shown that the behavior of the perturbations is damped, forward in time, by these new features with respect to the standard case of an ideal fluid. In the present work we generalize the investigation to the whole class of quasi-isotropic solutions. Our aim is to determine the conditions on the parameter $\zeta_0$ (\emph{i.e.}, on the viscosity intensity) which allows for the existence of a quasi-isotropic regime for the radiation dominated Universe.

For this purpose, we investigate the Einstein equations under the assumptions proper of the quasi-isotropic model. We separate zeroth- and first-order terms into the 3-metric tensor and the whole analysis follows this scheme of approximation. In the search for a self-consistent solution, we make use of the hydrodynamics equations in view of fixing the form of the energy density. As a result, we prove the existence of a quasi-isotropic solution, which has a structure analogous that provided by Khalatnikov, Kamenshchik and Starobinsky \cite{kks02}. Of course, in our solution the power-law for the leading 3-metric term is sensitive to the viscosity parameter $\zeta_0$. In particular, we find that such a solution exists only if when $\zeta_0$ remains smaller than a certain critical value. Finally, we determine the density contrast and its dependence on $\zeta_0$. This behavior confirms and generalizes the result obtained in \cite{nkm05} about the damping of density perturbations by the viscous correction.

The scheme of the paper is as follows. In Sec. 2, a review of the quasi-isotropic solution \cite{lk63} is presented; after setting the form of the 3-metric expansion, the integration of the Einstein equations is performed in order to obtain a solution for the energy density, the density contrasts and the 3-velocity of the perfect fluid filling the space time. In Sec. 3, we write down the expression of the 3-metric expansion in a more general form without fixing the time power-law exponents. The construction of the generalized Ricci tensor follows. In Sec. 4, we introduce the viscous-matter contribution into the Einstein equations and then the energy-momentum tensor is written down in presence of bulk viscosity. Furthermore, the Taylor expansion of the energy density is addressed using the $00$-components of the gravitational equations. The viscous parameter $s$ is here set equal to $\nicefrac{1}{2}$ and a brief analysis of its physical meaning is performed. In Sec. 5, some comments on the validity of the addressed model are presented. We firstly discuss the role of the microphysical horizon concerning the thermal equilibrium of the expanding Universe and the related hydrodynamical treatment. Then we analyze in some details the effects of the first- (shear) viscosity and show how it must be neglected in the dynamics, for the appearance of internal consistences in the model. Finally, a comment on the introduction of a causal thermodynamics into the very early dynamics is addressed. In particular, we treat the viscous fluids description by using a \emph{relaxation-equation}. In Sec. 6, we integrate the $00$-Einstein equation matched together with the hydrodynamical one. The integration is faced separately for the zeroth- and first-order analysis and a critical threshold for the viscous parameter $\zeta_0$ is found by imposing the constraint for the consistence of the adopted model. As a result, we get the expressions of the energy density and the density contrast in presence of bulk viscosity. In Sec. 7, we integrate the $0\alpha$- and $\alpha\beta$-gravitational equations and an expression of the 3-velocity as a function of the perturbed metric term is found. Brief concluding remarks follow.

\section{Review on the LK-quasi-isotropic solution}
In 1963, E.M. Lifshitz and I.M. Khalatnikov \cite{lk63} first proposed the so-called quasi-isotropic solution. This model is based on the idea that the space contracts maintaining linear-distance changes with the same time dependence order by order (\emph{i.e.}, a Taylor expansion of the 3-metric is addressed). In this approach, the Friedmann solution becomes a particular case of a larger class of solutions existing only for space filled with matter \cite{lk64}.

The metric evolution is strongly characterized by the matter equation of state. For an ultra-relativistic perfect fluid, characterized by the relation $p=\epsilon/3$, the spatial metric assumes the form $\gamma_{\alpha\beta}\sim a_{\alpha\beta}\,t$, asymptotically as $t\rightarrow0$ (the cosmological singularity is set by convention in $t=0$), where $a_{\alpha\beta}$ are assigned functions of the coordinates. As a function of time, the 3-metric is expandable in powers of $t$. The quasi-isotropic solution is formulated in a synchronous system (\emph{i.e.}, $g_{0\alpha}=0$, $g_{00}=-1$), which is not strictly a co-moving one. The line element writes as
\begin{equation}
ds^2 = - dt^2 + \gamma_{\alpha\beta}(t, x^\gamma)dx^{\alpha}dx^{\beta}\;,
\end{equation}
with a spatial metric of the form
\begin{equation}
\label{metric-lk}
\gamma_{\alpha\beta}=t\;{a}_{\alpha\beta}+t^{2}\;{b}_{\alpha\beta}+...\;,\qquad
\gamma^{\alpha\beta}=t^{-1}\;a^{\alpha\beta}-\;{b}^{\alpha\beta}\;,
\end{equation}
where $a^{\alpha\beta}$ is defined as $a^{\alpha\beta}a_{\beta\gamma}=\delta_\gamma^\alpha$; furthermore, the relation $b_\beta^\alpha=a^{\alpha\gamma}b_{\gamma\beta}$ is ensured by the scheme of approximation.

The Einstein equations in the synchronous system assume the form \cite{ll-field}
\begin{subequations}
\label{e-equations}
\begin{align}
\label{e-equations1}
&R^0_0 = \tfrac{1}{2}\,\partial_t \kappa_{\alpha}^{\alpha} +
\tfrac{1}{4}\,\kappa_{\alpha}^{\beta}\kappa_{\beta}^{\alpha} = \;T^0_0-\tfrac{1}{2}T \;,\\
\label{e-equations2}
&R^0_\alpha = \tfrac{1}{2}\,(\kappa^{\beta}_{\beta ;\,\alpha}-
\kappa^{\beta}_{\alpha ;\,\beta}) = \; T_\alpha^0 \;,\\
\label{e-equations3}
&R^\beta_\alpha = \tfrac{1}{2\sqrt{\gamma}}\,\partial_t(\sqrt{\gamma}\,
\kappa_{\alpha}^{\beta}) + P_{\alpha }^{\beta } = \;T_\alpha^\beta-\tfrac{1}{2}T \delta_\alpha^\beta\;,
\end{align}
\end{subequations}
where $c$=$G$=$1$ and the tensor $\kappa_{\alpha\beta}$ and its contractions read
\begin{equation}
\kappa_{\alpha\beta} =\p_t\gamma_{\alpha\beta} =a_{\alpha\beta} +
2\,t\; b_{\alpha\beta}\;,\quad
\kappa^\beta_\alpha  =\gamma^{\beta\delta}\,\kappa_{\alpha\delta} = t^{-1}\;\delta^\beta_\alpha + 
\;b^\beta_\alpha\;,\quad
\kappa\;=\p_t\ln \sqrt{\gamma} = 3\,t^{-1}\, +\, b\;,
\end{equation}
and $\gamma=\det(\gamma_{\alpha\beta})\sim t^3(1+tb)\det(a_{\alpha\beta})$. Matter is described by an ultra-relativistic perfect fluid energy-momentum tensor
\begin{equation}
\label{en-mom-tensor0}
T_{\mu\nu}=(p+\epsilon)\,u_{\mu}u_{\nu}+p\,g_{\mu\nu}=
\frac{\epsilon}{3}\;(4u_{\mu}u_{\nu}+g_{\mu\nu})\;,
\end{equation}
which provides the following identities
\begin{equation}
T_0^0=\tfrac{1}{3}\,\epsilon\,(-4u_{0}^2+1)\;,\quad
T_\alpha^0=\tfrac{4}{3}\,\epsilon\,u_{\alpha}u^0\;,\quad
T_\alpha^\beta=-\tfrac{4}{3}\,\epsilon\,u_{\alpha}u^\beta\;,\quad T=0\;.
\end{equation}

Calculating the left-hand side of \reff{e-equations1}, \reff{e-equations2} up to zeroth- $\mathcal{O}(1/t^2)$ and first-order $\mathcal{O}(1/t)$ in $1/t$, we rewrite them respectively as
\begin{equation}
\label{4.3-4.4}
-\frac{3}{4\,t^{2}}+\frac{b}{2\,t}=\frac{\epsilon}{3}\;(-4u_{0}^2+1)\;,\qquad
\frac{1}{2}\,(b_{;\,\alpha}-b^{\beta}_{\alpha ;\,\beta})=
-\frac{4\,\epsilon}{3}\;u_{\alpha}u_0\;.
\end{equation}
Because of the identity $-1=u_\mu u^\mu\sim-u_0^2+t^{-1}u_\alpha u_\beta\,a^{\alpha\beta}$, it is immediate to see that $\epsilon\sim t^{-2}$ and $u_\alpha\sim t^{2}$; hence, in the asymptotic limit $t\rightarrow0$, $u_0^2\simeq1$ $(u_0=-1)$. From the first equation of \reff{4.3-4.4}, one can find the first two terms of the energy density expansion, while, from the second equation, the leading term of the velocity arises
\begin{equation}
\label{e-u}
\epsilon=\frac{3}{4\,t^{2}}\,-\,\frac{b}{2\,t}\;,\qquad	
u_\alpha=\frac{t^{2}}{2}\,(b_{;\,\alpha}-b^{\beta}_{\alpha ;\,\beta})\;.	
\end{equation}
Because of \reff{e-u}, the expression for the density contrast $\delta$ can be found as first and zeroth-order energy density ratio, \emph{i.e.},
\begin{equation}
\delta\,=\,-\tfrac{2}{3}\,b\;t\;.
\end{equation}
This behavior implies that, as expected in the cosmological standard model, the zeroth-order term of energy density diverges more rapidly than the perturbations and the singularity is naturally approached with a vanishing density contrast in this scenario.

Besides the solutions for $\epsilon$ and $u_\alpha$, one has to consider the pure spatial components of the gravitational equation \reff{e-equations3}. Up to first approximation, the Ricci tensor can be written as $P_\alpha^\beta=\tilde{P}_\alpha^\beta/t$, where $\tilde{P}_\alpha^\beta$ is constructed by the constant 3-tensor $a_{\alpha\beta}$. The terms of order $t^{-2}$ automatically cancel out, while those proportional to $t^{-1}$ give
\begin{equation}
\label{eq-u}
\tilde{P}_{\alpha}^{\beta}\;+\;\tfrac{3}{4}\,b_{\alpha}^{\beta}\;+\;
\tfrac{5}{12}\,b\,\delta_{\alpha}^{\beta}=0\;.
\end{equation}
Performing the trace of this equation, a relation between the quite arbitrary six functions $a_{\alpha\beta}$ and the coefficients $b_{\alpha\beta}$ from the next-to-leading term of expansion can be determined: $b_{\alpha}^{\beta}=-\nicefrac{4}{3}\tilde{P}_{\alpha}^{\beta}+ \nicefrac{5}{18}\,\tilde{P}\,\delta_{\alpha}^{\beta}$. It is worth reminding that, in the asymptotic limit $t\rightarrow0$, the matter distribution becomes homogeneous because $\epsilon$ approaches a value independent of $b$. 

Now, using the Ricci identity $\tilde{P}^{\beta}_{\alpha;\,\beta}=\nicefrac{1}{2}\,\tilde{P}_{\,;\,\beta}$, the useful relation $b^{\beta}_{\alpha ;\,\beta}=\nicefrac{7}{9}\,b_{;\,\alpha}$ can be determined; this gives the final expression for the 3-velocity distribution as
\begin{equation}
\label{u}
u_\alpha=\frac{t^{2}}{9}\;b_{;\,\alpha}\;.	
\end{equation}
This result implies that, in this approximation, the 3-velocity is a gradient field of a scalar function fixed by the non perturbed metric $a_{\alpha\beta}$. As a consequence, the curl of the velocity vanishes and no rotations take place into the fluid.

Finally, it must be observed that metric \reff{metric-lk} allows for an arbitrary 3-space coordinate transformation and the solution above contains only $6-3=3$ arbitrary space functions arising from $a_{\alpha\beta}$. A particular choice of this functions, those which correspond to the space of constant curvature ($\tilde{P}^{\beta}_{\alpha}\sim\delta^{\beta}_{\alpha}$), can reproduce the pure isotropic and homogeneous model.

\section{Generalized quasi-isotropic line element}
In order to generalize the quasi-isotropic solution of the Einstein equations for the presence of dissipative effects into the evolution of the energy source, we deal with a more complex (no longer in integer powers) form of the 3-metric \reff{metric-lk} \cite{mont99,mont00,kks02}. In this respect, we take the spatial metric as
\begin{equation}
\gamma_{\alpha\beta}=t^{x}\;{a}_{\alpha\beta}+t^{y}\;{b}_{\alpha\beta}\;,\qquad\quad
\gamma^{\alpha\beta}=t^{-x}\;{a}^{\alpha\beta}-t^{y-2x}\;{b}^{\alpha\beta}\;.
\end{equation}
Here, the constraints for the space contraction (\emph{i.e.}, $x>0$), and for the consistence of the perturbation scheme (\emph{i.e.}, $y>x$) have to be imposed for the proper development of the model. In this approach, the extrinsic curvature and its contractions read
\begin{subequations}
\begin{align}
&\kappa_{\alpha\beta} =x\,t^{x-1}\; a_{\alpha\beta} +
y\,t^{y-1}\; b_{\alpha\beta}\;,\\
&\kappa^\beta_\alpha  = x\,t^{-1}\;\delta^\beta_\alpha + 
(y-x)\,t^{y-x-1}\;b^\beta_\alpha\;,\\
&\kappa\;= 3x\,t^{-1}\; +(y-x)\,t^{y-x-1}\; b\;,
\end{align}
\end{subequations}
furthermore, we calculate the following useful relation
\begin{equation}
\p_t\ln\sqrt{\gamma} =\tfrac{1}{2}\;\kappa =\tfrac{3}{2}\,xt^{-1}+
\tfrac{1}{2}\,(y-x)\,t^{y-x-1}\; b\;.
\end{equation}

We are now able to write down the final form of the Ricci-tensor components contained in the Einstein equations \reff{e-equations}. These new expressions allow us to generalize the original quasi-isotropic approach. Our aim is to obtain constraints and relations for the exponents $x$, $y$ in order to guarantee the existence of the solutions of our model. They explicitly read 
\begin{subequations}
\begin{align}
&R_0^0 = -\frac{3x(2-x)}{4t^2}\,+\,(y-x)(y-1)\frac{b}{2t^{2-y+x}}\;,\\
&R_{\alpha}^0 = (b_{\,;\,\alpha}-b_{\alpha;\,\beta}^{\beta})\frac{y-x}{2t^{1-y+x}}\;,\\
&R_{\alpha}^{\beta} = \frac{x(3x-2)}{4t^2}\,\delta_{\alpha}^{\beta}\,+\,
\frac{(y-x)(2y+x-2)}{4t^{2-y+x}}\,b_{\alpha}^{\beta}\,+\;
\frac{(y-x)x}{4t^{2-y+x}}\,b\,\delta_{\alpha}^{\beta}\,+\,
\frac{\tilde{P}_{\alpha}^{\beta}}{t^{x}}\,+\,
\frac{P_{\alpha}^{*\beta}}{t^{2x-y}} \;.\label{R-ab}
\end{align}
\end{subequations}
We note that in equation \reff{R-ab}, $\tilde{P}_{\alpha}^{\beta}$ represents the 3-dimensional Ricci tensor constructed by the metric $a_{\alpha\beta}$. On the other hand, the higher-order term $P_{\alpha}^{*\beta}$ denotes the part of $P_{\alpha}^{\beta}$ containing the 3-tensor $b_{\alpha\beta}$.

\section{The form of energy density in the viscous approach}
In the quasi-isotropic solution the Universe is assumed, in correspondence with the standard cosmological model, to be described by the energy-momentum tensor of an ultra-relativistic perfect fluid. In connection with the development of new cosmological models, the discovery of the cosmic acceleration suggests matter to play an essential role at different stages of cosmological evolution and it can obey very different equations of state \cite{ss00}. Thus, corrections in this sense to the original formulation of the quasi-isotropic solution can be useful in this new context. 

In this work, we treat the immediate generalization of LK scheme. We consider the presence of dissipative processes within the fluid dynamics, as expected at temperatures above $\mathcal{O}(10^{16} GeV)$; this extension is represented by an additional term in the expression of the energy-momentum tensor \reff{en-mom-tensor0} and it can be derived from thermodynamical properties of the fluid \cite{nkm05,clw92,mont01}. The new tensor reads
\begin{equation}
\label{en-mom-tensor}
T_{\mu\nu}=(\tilde{p}+\epsilon)u_{\mu}u_{\nu}+\tilde{p}\,g_{\mu\nu}=
\frac{\epsilon}{3}\;(4u_{\mu}u_{\nu}+g_{\mu\nu})-
\zeta\,u^{\rho}_{;\,\rho}(u_{\mu}u_{\nu}+g_{\mu\nu})\:,
\end{equation}
\begin{equation}\label{p-tot}
\tilde{p}= p - \zeta\,u^{\rho}_{;\,\rho}\;,
\end{equation}
where $p=\epsilon/3$ denotes the usual thermostatic pressure in correspondence of an ultra-relativistic state equation and $\zeta$ is the bulk viscosity coefficient. In this work, we neglect the so-called shear viscosity (first viscosity) since, in the case of quasi-isotropic cosmological evolution, the displacement of the matter layers with respect to each other appears only in a higher-order analysis (this kind of viscosity represents the energy dissipation due to such an effect). 

The coefficient $\zeta$ has to be expressed in terms of the thermodynamical parameters of the fluid. In particular, here we assume this quantity as a function of the Universe energy density; according to literature developments \cite{bk77,b87,b88}, we express $\zeta$ as a power-law of the form
\begin{equation}
\label{bulk}
\zeta=\zeta_0\,\epsilon^s\;,	
\end{equation}
where $\zeta_0$ is a constant and $s$ is a dimensionless parameter whose behavior was discussed by V.A. Belinskii et al. \cite{bk76,bnk79} for asymptotic values of the energy density. This study yields the constraint $0\leqslant s\leqslant\nicefrac{1}{2}$ for its range of variation in correspondence of large values of $\epsilon$.

Let us now write the expressions of the mixed components of tensor \reff{en-mom-tensor} up to higher-order corrections as 
\begin{subequations}
\begin{align}
\label{T00}
&T^0_0 = -\frac{\epsilon}{3}\,(4u_0^2-1)+\zeta_0\epsilon^s
\,u^{\mu}_{;\,\mu}\,(u_0^2-1)\;,\\
\label{TTTT}
&T^{}_{\,}\; = -3\,\zeta_0\,\epsilon^s\,u^{\mu}_{;\,\mu}\;,\\
&T^\beta_\alpha = \frac{\epsilon}{3}\,(4u_\alpha u^\beta+\delta_\alpha^\beta)-
\zeta_0\epsilon^s\,u^{\mu}_{;\,\mu}\,(u_\alpha u^\beta+\delta_\alpha^\beta)\;,\\
&T^0_\alpha = \tfrac{4}{3}\,\epsilon\,u_\alpha u^0-
\zeta_0\epsilon^s\,u^{\mu}_{;\,\mu}\;u_\alpha u^0\;,
\end{align}
\end{subequations}
where the divergence of the 4-velocity reads as
\begin{equation}\label{4-divergence}
u^{\mu}_{;\,\mu}=\p_t\ln\sqrt{\gamma}=
\tfrac{3}{2}\,xt^{-1}+\tfrac{1}{2}\,(y-x)\,t^{y-x-1}\; b\;.
\end{equation}
Here, we assume, as in the non-viscous case, the relation: $u_0^2\simeq1$ (with $u_0=-1$), whose consistence must be verified \emph{a posteriori} comparing the time behavior of the quantities involved in the model. Taking into account expressions \reff{T00} \reff{TTTT}, we can recast now the Einstein equation \reff{e-equations1} in the form
\begin{equation}
\label{o-o}
-\frac{3x(2-x)}{4t^2}\,+\,(y-x)(y-1)\frac{b}{2t^{2-y+x}}\,=\,
-\epsilon\,+\,\frac{9\,x}{4\,t}\;\zeta_0\epsilon^s\,+\,
\frac{3(y-x)}{4\,t^{1-y+x}}\;\zeta_0\epsilon^s\,b\;.
\end{equation}

In what follows, we fix the value $s=\nicefrac{1}{2}$ in order to deal with the maximum effect that bulk viscosity can have without dominating the dynamics. In fact, the notion of this kind of viscosity corresponds to a phenomenological issue of perturbations to the thermodynamical equilibrium \cite{bk77,nkm05}. In this sense, we remark that, if $s>\nicefrac{1}{2}$, the dissipative effects become dominant and non-perturbative. Moreover, if we assume the viscous parameter $s<\nicefrac{1}{2}$, the dynamics of the early Universe is characterized by an expansion via a power-law $a(t)\sim t^{2/3\gamma}$ starting from a perfect fluid Friedmann singularity at $t=0$ (here $\gamma$ is identify by the relation $p=(\gamma-1)\,\epsilon$). After this first stage of evolution, where viscosity does not affect at all the dynamics, the Universe inflates in the limit $t\rightarrow\infty$ (\emph{i.e.}, out of our approximation scheme) to a viscous deSitter solution characterized by $a(t)\sim e^{H_0t}$, $H_0=\sqrt{\epsilon_0}/3=\nicefrac{1}{3}\,\, (\zeta_0\sqrt{3}/\gamma)^{1/(1-2s)}$ \cite{b88,b90}.

Since, in this work, we deal with the asymptotic limit $t\rightarrow0$, we only treat the case $s=\nicefrac{1}{2}$ in order to quantitatively include dissipative effects in the primordial dynamics. From equation \reff{o-o}, if $s=\nicefrac{1}{2}$, we expand the energy density $\epsilon$ as follows
\begin{equation}
\label{epsilon-}
\epsilon=\frac{e_0}{t^2}\;+\;\frac{e_1\,b}{t^{2-y+x}}\;,\qquad
\sqrt{\epsilon}=\frac{\sqrt{e_0}}{t}\left(1\;+\;\frac{e_1\,b}{2e_0}\,t^{y-x}\right)\;,
\end{equation}
where the constants $e_0$, $e_1$ are to be determined combining the $00$-gravitational equation with the hydrodynamical ones, comparing the terms order by order, as treated below. We remark that only for the case $s=\nicefrac{1}{2}$ all terms coming on the left-hand and the right-hand side respectively of equation \reff{o-o} result to have the same time behavior up to first order because of \reff{epsilon-}.

\section{Comments on the adopted paradigm}
In this Section, we discuss in some details the hypotheses at the ground of our analysis of the quasi-isotropic viscous Universe dynamics. In particular, we investigated some peculiar features of the very early evolution (near the cosmological singularity) since their presence leads to a specific treatment of the viscous phenomena.

It is well known \cite{kolb} the crucial role played in cosmology by the \emph{microphysical horizon}, as far as the thermodynamical equilibrium is concerned. In the isotropic Universe, such a quantity is fixed by the inverse of the expansion rate, $H^{-1}\equiv (a/\dot{a})$ ($a$ being the scale factor of the Universe and the dot identifies time derivatives) and it gives the characteristic scale below which the elementary-particle interactions are able to preserve the thermal equilibrium of the system. Therefore, if the mean free-path of particles $\ell$ is greater than the microphysical horizon (\emph{i.e.}, $\ell>H^{-1}$), no real notion of thermal equilibrium can be recovered at the micro-causal scale. If we indicate by $n$ the number density of particles and by $\sigma$ the averaged cross section of interactions, then the mean free-path of the ultra-relativistic cosmological fluid (in the early Universe the particles velocity is very close to speed of light) takes the form 
$\ell\sim 1/n\sigma$. Interactions mediated by massless gauge bosons are characterized by the cross section $\sigma\sim\alpha^{2}\,T^{-2}$ ($g=\sqrt{4\pi\alpha}$ being the gauge coupling strength) and the physical estimation $n\sim T^{3}$ leads to the result $\ell\sim1/\alpha^{2}T$ \cite{kolb}. During the radiation-dominated era $H\sim T^{2}/m_{Pl}$, so that
\begin{equation}
\ell\sim \frac{T}{\alpha^2m_{Pl}}\;H^{-1}\;.
\end{equation}
Therefore in the case of $T\apprge\alpha^2m_{Pl}\sim \mathcal{O}(10^{16}GeV)$, \emph{i.e.}, during the earliest epoch of pre-inflating Universe, the interactions above are effectively ``frozen out'' and they are not able to maintain or to establish thermal equilibrium. To complete this consideration we remark that, at temperatures grater than $\mathcal{O}(10^{16}GeV)$, the contributions to the estimation above due to the mass term of the gauge bosons can be ruled out for all known and proposed perturbative interactions.

As a consequence of this non-equilibrium configuration of the causal regions characterizing the early Universe, most of the well-established results about the kinetic theory \cite{ll-pk,weinberg,weinberg71} concerning the cosmological fluid nearby equilibrium become not applicable. Indeed all these analysis are based on the assumption to deal with a finite mean free-path of the particles and, in particular, results about the characterization of viscosity are established when pure collisions among particles are retained. However, when the mean free-path is grater than the micro-causal horizon (which, in the pre-inflating Universe, coincides with the cosmological horizon), $\ell$ can be taken of infinite magnitude for any physical purpose.

The fundamental analysis of the viscous cosmology is due to the Landau school \cite{bk76,bk77,bnk79}; since they were aware of these difficulties for a consistent kinetic theory, such an analysis was essentially based on an hydrodynamical approach. A real notion of the \emph{hydrodynamical description} can be provided by assuming that an arbitrary state is adequately specified by the particle-flow vector and the energy momentum tensor alone \cite{is79}. In particular, the entropy flux has to be expressible as a function of these two hydrodynamical variables without additional parameters. Following this point of view, the viscosity effects are treated on the ground of a thermodynamical description of the fluid, \emph{i.e.}, the viscosity coefficients are fixed by the macroscopic parameters which govern the system evolution. In this respect, the most natural choice is to take such a (shear and bulk) viscosity coefficients as power-laws in the energy density of the fluid (for a detailed discussions see \cite{bk76}). Such a phenomenological assumption can be reconciled, for some simple cases, with a relativistic kinetic theory approach \cite{murphy73}, especially in the limits of small and large energy densities.  

Addressing the hydrodynamical approach, we are lead to retain the same equation of state which would characterize the corresponding ideal fluid. This fact is supported by idea that the viscosity effects provide only small corrections to the thermodynamical setting of the system. As clarified above, in the present analysis, we deal with the case in which bulk viscous corrections are of the same order of the perfect fluid contributions, in order to maximize their influence in the Universe dynamics. Nevertheless, since we are treating an ultra-relativistic thermodynamical system, which is very weakly interacting on the micro-causal scale, there are well-grounded reasons to describe it by the equation of state $p=\epsilon/3$. 

Another important point concerning the ground assumptions of our model, is why the \emph{shear viscosity} ($\eta$) is not addressed in the present scheme. Indeed, this kind of viscosity accounts for the friction forces acting between different portions of viscous fluid. Therefore, as far as the isotropic character of the Universe is retained, the shear viscosity must not provide any contributions, as discussed in \cite{bk77}. On the contrary, the rapid expansion of the early Universe suggests that an important contribution comes out from the bulk viscosity as an averaged effect of a quasi-equilibrium evolution.

Indeed, our present analysis deals with small inhomogeneous corrections to the background FRW-metric. Thus, at first-order in our solution, shear viscosity should be, in principle, included into the dynamics. In this sense it is shown in \cite{bk76} that, if the bulk viscosity coefficient behaves like $\zeta\sim\epsilon^{s}$, then the corresponding shear one behaves as $\eta\sim\epsilon^{p}$, where $p$ must satisfy the constraint condition $p\geqslant s+\nicefrac{1}{2}$. Here we treat the case $s=\nicefrac{1}{2}$, thus getting $p\geqslant1$ for the $\eta$ coefficient. This issue is incompatible with the symmetries and the approximations here addressed. In fact, the shear viscosity provides, among others, an equivalent contribution to the bulk one, since the energy-momentum tensor of the viscous fluid contains the term
\begin{equation}
T_{\mu\nu}\sim\,...\,-\,(\zeta-\tfrac{2}{3}\eta)\,u^{\rho}_{;\,\rho}
(u_{\mu}u_{\nu}+g_{\mu\nu})\,+\,...\;.
\end{equation}

We now observe that, at zeroth-order, $u^{\rho}_{;\,\rho}\sim\mathcal{O}(1/t)$, while the first-order correction in the energy density behaves like $\mathcal{O}(1/t^{x})$ and we will show the relation $1\leqslant x<2$ in Sec. 6. Since the request $x\geqslant1$ comes out from the zeroth-order analysis, which by isotropy is independent of the shear contribution, we can conclude that, for our model, the shear viscosity would produce the inconsistency associated to the term $\mathcal{O}(1/t^{px+1})$. The point is that the request $px+1\geqslant2$ would make such a contribution dominant in the model, against the basic assumption. Thus, to include shear viscosity in a quasi-isotropic model, we should choose the case $s<\nicefrac{1}{2}$ which is out of the aim of this paper since it is devoted to maximize the bulk effects in a coherent cosmological dynamics.

To conclude this section, we would like to discuss the question concerning the implementation of a causal thermodynamics for our cosmological model. Indeed, the hydrodynamical theory of a viscous fluid is applicable only when the spatial and temporal derivatives of the velocity of the matter are small \cite{iv70,israel76,is79}. This condition is necessarily violated in the asymptotic limit near the cosmological singularity. This way viscous fluids would have to be described by using a relaxation equation similar to the Maxwell equation in the theory of viscoelasticity \cite{bnk79}.

In  this scheme, the energy-momentum tensor assumes the form
\begin{equation}
\label{sigma-en-mom-tensor}
T_{\mu\nu}=\epsilon\,u_{\mu}u_{\nu}-(p+\sigma)\,(g_{\mu\nu}+u_{\mu}u_{\nu})\;,
\end{equation}
where $p$ denotes the thermostatic pressure and $\sigma$ is the \emph{bulk-stress density}. In the very early Universe, the relation between $\sigma$ and the relaxation time $\tau_0$ reads as follow
\begin{equation}
\label{sigma}
\sigma\;+\;\dot{\sigma}\,\tau_0	\;=\;\zeta\,u^{\rho}_{;\,\rho}\;.
\end{equation}
The relaxation time can be expressed as $\tau_0/\zeta\sim1/\epsilon\;$: this physical assumption follows from the fact that the transverse-wave velocity in matter has finite (nonzero) magnitude in the case of large values of $\epsilon$ \cite{bnk79}. 

In this scheme, we are able to express the time dependence of $\tau_0$. Since, at leading order, $\epsilon\sim1/t^{2}$, we obtain, using Eq. \reff{bulk}, the following behavior for the relaxation time $\tau_0\sim t^{2-2s}$. In our model we deal with the case $s=\nicefrac{1}{2}$ which yields $\tau_0\sim t$ and, if we address a power-law dependence on $\sigma$ (according the structure of the solution) such as $\dot{\sigma}\sim \sigma/t$ \cite{bnk79}, relation \reff{sigma} rewrites as
\begin{equation}
\sigma=\tilde{\zeta}_0\,\epsilon^{s}\;u^{\rho}_{;\,\rho}\;.
\end{equation}
From this analysis we can apply the standard expression for the bulk viscous hydrodynamic taking into account the reparameterization $\zeta_0\rightarrow\tilde{\zeta}_0$ of the bulk coefficient.

The considerations above allow us to regard the subtle paradigm of the causal thermodynamics, having in mind that it would affect only qualitative details of our analysis, but it could not alter the validity of our results.

\section{Solutions of the 00-Einstein and the hydrodynamical equations}
In Section 4, we exploited Eq. \reff{e-equations1} in order to obtain the qualitative expression for the energy density $\epsilon$ when the matter filling the space is described by a viscous fluid energy-momentum tensor. We now match equation \reff{o-o} rewritten as   
\begin{align}
\label{r00}
&\big[-\tfrac{3}{4}\,x\,(2-x)+\,e_0\,-\tfrac{9}{4}\,\zeta_0\,x\,\sqrt{e_0}\;\big]
\;t^{-2}+\nonumber\\
&+\big[\tfrac{1}{2}(y-x)(y-1)+\,e_1\,-\tfrac{9}{8}\,\zeta_0\,x\,e_1\,e_0^{-1/2}-
\tfrac{3}{4}(y-x)\,\zeta_0\,\sqrt{e_0}\big]\;b\,t^{y-x-2}=0\;,
\end{align}
with the hydrodynamical ones $T_{\mu;\,\nu}^{\nu}=0$. It is worth noting that, in the non-viscous case ($\zeta_0=0$), the energy density solution is determined without exploiting the hydrodynamical equations, as in \cite{kks02}, since $\epsilon$ directly comes out from the $00$-gravitational equation. In our approximation ($u_\alpha$ is neglected with respect to $u_0$), the energy-momentum tensor conservation law provides the equation
\begin{equation}
\p_t\epsilon+\p_t(\ln\sqrt{\gamma})\,\left[\tfrac{4}{3}\epsilon-\zeta_0\epsilon^s\p_t(\ln\sqrt{\gamma})\right]=0\;,
\end{equation}
which can be simplified as follows
\begin{align}
\label{hydro}
\big[2e_0(x-1)&-\tfrac{9}{4}\,\zeta_0\,x^{2}\sqrt{e_0}\,\big]\;t^{-3}+\nonumber\\
+&\big[e_1\big(b(y-x-2)+2xb-\tfrac{9}{8}\,\zeta_0 x^{2}b\,e_0^{-1/2}\;\big)+\nonumber\\
&\qquad\qquad+\tfrac{2}{3}(y-x)\,b\,e_0-\tfrac{3}{2}\,x(y-x)\,\zeta_0\,b\sqrt{e_0}\,\big]
\;t^{y-x-3}=0\;.
\end{align}

Equations \reff{r00} and \reff{hydro} have to be combined together and solved order by order in the expansion in $1/t$ (in the asymptotic limit $t\rightarrow0$). Since for the coherence of the solution, we impose $y>x$, by solving the leading-order identities we get
\begin{equation}
\label{0-sol}
x=\frac{1}{1-\tfrac{3\sqrt{3}}{4}\,\zeta_0}\;,\qquad\quad 
e_0=\tfrac{3}{4}\;x^{2}\;.
\end{equation}
The parameter $\zeta_0$ has here the restriction $\zeta_0\leqslant\nicefrac{4}{3\sqrt{3}}\,$ in order to satisfy the condition $x>0$. In this way the exponent of the metric power-law $x$ runs from $1$ (which corresponds to the non viscous limit $\zeta_0=0$) to $\infty$ \cite{nkm05,kks02}. We remark that this constraint on $\zeta_0$ arises from a zeroth-order analysis and defines the existence of a viscous Friedmann-like model, in which the early Universe has to expand with positive powers of time.  

Comparing now the two first-order identities (which involve the terms pro-portional to $t^{y-x-2}$ and $t^{y-x-3}$), we easily get an algebraic equation for the $y$ parameter
\begin{equation}
y^{2}-y(x+1)+2x-2=0\;.
\end{equation}
The solutions are $y=2$, $y=x-1$\footnote{We remark that in \cite{kks02} (see Eq.(34), (35)) this solution is found by imposing the consistence of the $\alpha\beta$-Einstein equation and not as a pure dynamical condition derived by the solution of the perturbed hydrodynamical equation.}. Obviously the second one does not respect the condition $y>x$; hence the first order correction to the 3-metric is characterized by the following values
\begin{equation}
y=2\;,\qquad\quad 
e_1=-\tfrac{1}{2}\,x^{3}+2x^{2}-2x\;.
\end{equation}
It is immediate to see that, in the non viscous case $\zeta_0=0$, we obtain $x=1$, $e_0=\nicefrac{3}{4}$, $e_1=-\nicefrac{1}{2}$, which reproduce the energy density solution \reff{e-u}. 

By guaranteeing the consistence of the model, we now narrow the validity of the parameter $x$ to the values which satisfy the constraint $x<y$. Thus, from \reff{0-sol}, the quasi-isotropic solution exists only if 
\begin{equation}
\label{zzero}
\zeta_0<\zeta_0^{*}=\frac{2}{3\sqrt{3}}\;,
\end{equation}
\emph{i.e.}, the viscosity is sufficiently small. For values of the viscous parameter $\zeta_0$ that overcome the critical one ($\zeta_0^{*}$), the quasi-isotropic expansion in the asymptotic limit as $t\rightarrow0$ can not be addressed, since perturbations would grow more rapidly than the zeroth-order terms. It is worth noting that the study of the perturbation dynamics in a pure isotropic picture yields a very similar asymptotic behavior when viscous effects are taken into account \cite{nkm05,paddy}. The Friedmann-singularity scheme is preserved only if we deal with limited values of the viscosity parameter, in particular we obtain the condition $\zeta_0^{(iso)}<\zeta_0^{*}/3$: this constraint is physically motivated if we consider, as it is,  the Friedmann model as a particular case of the quasi-isotropic solution.

\emph{Comments on the total pressure sign:}\quad The solution of the unperturbed dynamics gives rise to the expression of the metric exponent $x$ in terms of the viscous parameter $\zeta_0$ and to the zeroth-order expression of the energy density, which reads
\begin{equation}
\epsilon=\frac{3x^{2}}{4t^{2}}+...\;.
\end{equation}
In order to characterize the effective expansion of the early Universe, let us now recall the expression of the total pressure $\tilde{p}$ \reff{p-tot} at leading-order:
\begin{equation}
\tilde{p}=\tfrac{1}{3}\;\epsilon+\tfrac{3}{2\,t}\;\zeta_0\;\sqrt{\epsilon}\;x\;,
\end{equation}
where we have used the 4-divergence \reff{4-divergence} truncated at zeroth-order. By using these identities, the condition $\tilde{p}\geqslant0$ yields the inequality
\begin{equation}
\zeta_0\leqslant\zeta_0^{*}/2\;,	
\end{equation}
which strengths the constraint \reff{zzero} and restricts the $x$-domain to $[1,\nicefrac{4}{3}]$.

The request to deal with a positive (at most zero) total pressure is consistent with the idea that bulk viscosity must not drastically change the standard dynamics of the isotropic Universe. In this respect, we address the domain $\zeta_0^{*}\leqslant\zeta_0^{*}/2$ as a physical restriction on the initial conditions for the existence of a well grounded quasi-isotropic solution.
\vspace{0.3cm}

By concluding this section, we rewrite the expression of the energy density in order to analyze the density contrast evolution. In presence of bulk viscosity, $\epsilon$ assumes the form
\begin{equation}
\label{e}
\epsilon=\frac{3\,x^{2}}{4\,t^2}\;-\;\frac{(x^{3}/2-2x^{2}+2x)\;b}{t^{x}}\;,
\end{equation}
and, hence, the density contrast $\delta$ can be written as
\begin{equation}
\delta\,=\,-\tfrac{8}{3}\,(x/4+1/x-1)\,b\;t^{2-x}\;.
\end{equation}
Since $x$ runs from $1$ to $2$ as the viscosity increases towards its critical value, we note that, as in \cite{nkm05}, the density contrast evolution is strongly damped by the presence of dissipative effects which act on the perturbations. In this sense, we remark that bulk viscosity can damp the evolution of perturbations forward in time. This behavior implies that the density contrast approaches the singularity, \emph{i.e.}, $\delta=0$, more weakly as $t\rightarrow0$ when the viscosity runs to $\zeta_0^{*}$. In correspondence with this threshold value the density contrast remains constant in time and hence it must be excluded by the possible $\zeta_0$ choices.

\section{The relation for the velocity and the 3-metric}
The $00$-Einstein equation provides the solution for the energy density; to perform a complete analysis of the quasi-isotropic model and to verify the consistence of our approximations, we now investigate the solutions of the $0\alpha$-components of the gravitational equations and the spatial $\alpha\beta$- ones.

Imposing the condition $s=\nicefrac{1}{2}$, the Einstein equation \reff{e-equations2} reads
\begin{equation}
\frac{y-x}{2\,t^{1-y+x}}\,\big(b_{\,;\,\alpha}-b_{\alpha;\,\beta}^{\beta}\big)=
\frac{4}{3}\,\epsilon\,u_\alpha - \zeta_0\sqrt{\epsilon}\, u_\alpha
\left(\frac{3x}{2t}+\frac{(y-x)b}{2\,t^{1-y+x}}\right)\;.
\end{equation}
Substituting \reff{e} in the last equation, we get the following expression for the velocity, up to the leading-order of expansion (here in particular we neglect terms of order $\mathcal{O}(t^{-1})$ and $\mathcal{O}(t^{1-x})$):
\begin{equation}
u_\alpha=\frac{2-x}{2x}\,(b_{,\alpha}-b^{\beta}_{\alpha ;\,\beta})\;t^{3-x}\;.
\end{equation}
It is worth noting that, in our generalization, the assumption $u_0^{2}\simeq1$ is well verified, since we immediately see that $u_\alpha u^\beta\sim t^{6-3x}$, which can be neglected in the 4-velocity contraction $u_\mu u^\mu=-1$; hence the approximated hydrodynamical equation \reff{hydro} is still self-consistent using this expression of $u_\alpha$.

Let us now write down equation \reff{e-equations3}: here, the first two leading-orders of the right-hand side are $\mathcal{O}(t^{-2})$ and $\mathcal{O}(t^{-x})$ respectively only if $x<2$ like in our scheme; hence $u_\alpha u^\beta$ is neglected, as seen before, $\mathcal{O}(t^{-2})$ terms cancel each other, while those proportional to $t^{-x}$ give the following equation (which generalize \reff{eq-u})
\begin{equation}
\label{eq-uu}
\tilde{P}_{\alpha}^{\beta}\;+\;A\,b_{\alpha}^{\beta}\;+\;
B\,b\,\delta_{\alpha}^{\beta}\;+\;C\,\delta_{\alpha}^{\beta}=0\;,
\end{equation}
where the quantities $A$, $B$, $C$ are defined as
\begin{equation}
A=\tfrac{1}{4}\,(4-x^{2})\,,\quad
B=\tfrac{1}{6}(2x-1)(x-2)^{2}-\tfrac{1}{4}\,x(x-2)\,,\quad
C=-\tfrac{1}{6}(2-x)(x-1)\;,
\end{equation}
respectively. Taking the trace of \reff{eq-uu}, we obtain the relation $(A+3B)\,b=-\tilde{P}\,-\,3C$ which provides the following equation
\begin{equation}
2A\;b^{\beta}_{\alpha;\,\beta}=(A+B)\,b_{,\alpha}\;,
\end{equation}
when combined with the Ricci 3-tensor relation
$\tilde{P}^{\beta}_{\alpha;\,\beta}=\nicefrac{1}{2}\tilde{P}_{\,;\,\beta}$.

Therefore we are now able to write down the final form of the 3-velocity related to the perturbed metric tensor trace $b$ :
\begin{equation}
u_\alpha=\frac{2-x}{4xA}\,(A-B)\;t^{3-x}\,b_{,\alpha}\;.
\end{equation}
As it can be easily checked, the solution here constructed matches the non-viscous one \reff{u} if we set $\zeta_0=0$ and it is completely self-consistent up to the first two orders in time. As in the original analysis, the present model contains only three physically-arbitrary functions of the spatial coordinates, \emph{i.e.}, the six functions $a_{\alpha\beta}$ minus three degrees of freedom ruled out by fixing suitable space coordinates. The only remaining free parameter of the model is viscous one, $\zeta_0$.

\section{Concluding remarks}
Our analysis outlined how the presence of bulk viscosity can deeply modify the quasi-isotropic solution in the asymptotic limit near cosmological singularity. The investigation started from the modification of the Einstein equations, induced by a viscous matter term and then proceeded by the integration of the new gravitational equations matched together with the hydrodynamical ones, order by order in the $1/t$ expansion.

As a main result, we have shown that the quasi-isotropic solution exists only for particular values of the bulk viscosity coefficient $\zeta_0$. When the dissipative effects become too relevant, we are not able to construct the solution following the line of the LK model. In fact, when $\zeta_0$ approaches the threshold value $\zeta_0^*=\nicefrac{2}{3\sqrt{3}}$, the approximation scheme fails and the model becomes non self-consistent.

By requiring that the viscosity parameter $\zeta_0$ be under its critical value, we have also outlined how the behavior of the density contrast is deeply influenced by the presence of bulk viscosity. In fact, as far as dissipative effects are taken into account, the density contrast contraction ($\delta\rightarrow0$ as $t\rightarrow0$), is damped until remaining constant if $\zeta_0$ assumes its critical value.

We conclude by stressing that our result is relevant near the singularity, where the volume of the quasi-isotropic Universe changes rapidly and as a consequence, the cosmological fluid has to follow this rapid variation by subsequent stages of thermal equilibrium. Then bulk viscosity emerges from the average non-equilibrium effects and it is expected to be increasingly relevant, when the singularity is approached.

\section*{Acknowledgments}

We would like to thank V.A. Belinskii for his valuable comments on these topics and for his help in dealing with the motivations for the adopted approach to describe viscous effects in the early Universe. We would also like to thank O.M. Lecian for her careful analysis of the manuscript and for the stylistic advice.


\begin{thebibliography}{0}
\newcommand{\mont}{G. Montani}

\newcommand{\cqg}{Class. Quant. Grav.}
\newcommand{\prl}{Phys. Rev. Lett.}
\newcommand{\pra}{Phys. Rev. A}
\newcommand{\prb}{Phys. Rev. B}
\newcommand{\prd}{Phys. Rev. D}
\newcommand{\apj}{Astrophys. J.}
\newcommand{\physrep}{Phys. Rep.}
\newcommand{\ptp}{Prog. Theore. Phys.}
\newcommand{\ijtp}{Int. J. Theore. Phys.}
\newcommand{\grg}{Gen. Rel. Grav.}
\newcommand{\pla}{Phys. Lett. A}
\newcommand{\plb}{Phys. Lett. B}
\newcommand{\pld}{Phys. Lett. D}
\newcommand{\ijmpa}{Int. J. Mod. Phys. A}
\newcommand{\ijmpb}{Int. J. Mod. Phys. B}
\newcommand{\ijmpd}{Int. J. Mod. Phys. D}
\newcommand{\mpla}{Mod. Phys. Lett. A}
\newcommand{\mplb}{Mod. Phys. Lett. B}
\newcommand{\advphys}{Adv. Phys.}
\newcommand{\sovJETP}{Sov. Phys. JETP}
\newcommand{\npb}{Nuc. Phys. B}
\newcommand{\JETPl}{JETP Lett.}


\newcommand{\bib}[6]{\bibitem{#1}{#2}, \emph{#3} \textbf{#4}, #5 (#6).}
\newcommand{\bibN}[7]{\bibitem{#1}{#2}, \emph{#3} \textbf{#4}{(#5)}, #6 (#7).}
\newcommand{\book}[5]{\bibitem{#1}{#2}, \emph{#3}, ({#4}, #5).}



\bib{lk63}{E.M. Lifshitz and I.M. Khalatnikov}{Adv. Phys.}{12}{185}{1963}
\bib{kks02}{I.M. Khalatnikov, Kamenshchik and Starobinsky}{\cqg}{19}{3845}{2002} 
\bibN{mont99}{\mont}{\cqg}{16}{3}{723}{1999}
\bibN{mont00}{\mont}{\cqg}{17}{11}{2197}{2000}
\bib{im03}{G.P. Imponente and G. Montani}{\ijmpd}{12}{1845}{2003}
\bibN{td94}{K. Tomita and N. Deruelle}{\prd}{50}{12}{7216}{1994}
\bibN{bk76}{V.A. Belinskii and I.M. Khalatnikov}{\sovJETP}{42}{2}{205}{1976}
\bibN{bk77}{V.A. Belinskii and I.M. Khalatnikov}{\sovJETP}{45}{1}{1}{1997}
\bibN{bnk79}{V.A. Belinskii, E.S. Nikomarov and Khalatnikov}{\sovJETP}{50}{2}{213}{1979}
\bib{b87}{J.D. Barrow}{\plb}{183}{285}{1987}
\bib{b88}{J.D. Barrow}{\npb}{310}{743}{1988}
\bibitem{b90}
J.D. Barrow, in \emph{The Formation and Evolution of Cosmic String}, eds. Gibbons G.W., Hawking S.W. and Vachaspati T. (Cambridge, 1990), p. 449.
\bibN{nkm05}{N. Carlevaro and G. Montani}{\mpla}{20}{23}{1729}{2005}
\bibN{paddy}{T. Padmanabhan and S.M. Chitre}{Phys. Lett. A}{120}{9}{433}{1987}
\bib{lk64}{E.M. Lifshitz and I.M. Khalatnikov}{Sov. Phys. Usp.}{6}{495}{1964} 
\book{ll-field}{L.D. Landau and E.M. Lifshitz}{The Classical Theory of Fields}{Pergamon Press}{1979}
\bib{ss00}{V. Sahni and A.A. Starobinsky}{\ijmpd}{9}{373}{2000}
\bib{clw92}{M.O. Calvao, J.A.S. Lima and I. Waga}{\pla}{162}{223}{1992}
\bib{mont01}{G. Montani}{\cqg}{18}{193}{2001}
\book{kolb}{E.W. Kolb and M.S. Turner}{The Early Universe}{Westview}{1994}
\book{ll-pk}{L.D. Landau and E.M. Lifshitz}{Physical Kinetics}{Pergamon Press}{1981}
\book{weinberg}{S. Weinberg}{Gravitation and Cosmology}{Wiley}{1972} 
\bib{weinberg71}{S. Weinberg}{\apj}{168}{175}{1971}
\bib{is79}{W. Israel and J.M. Stewart}{Ann. Phys.}{118}{241}{1979}
\bib{murphy73}{G. Murphy}{\prd}{8}{4231}{1973}
\bibN{iv70}{W. Israel and J.N. Vardalas}{Nuovo Cimento Letters}{IV}{19}{887}{1970}
\bib{israel76}{W. Israel}{Ann. Phys.}{100}{310}{1976}
\end{thebibliography}
\end{document}